\documentclass[twocolumn,preprintnumbers,amsmath,amssymb,superscriptaddress]{revtex4}
\usepackage{graphicx}
\usepackage{dcolumn}
\usepackage{bm}
\usepackage{soul}
\usepackage{color}
\usepackage{epstopdf}
\usepackage[version=3]{mhchem}
\usepackage{lipsum}
\usepackage[outercaption]{sidecap}
\usepackage{floatrow}
\usepackage{hyperref}
\usepackage{multirow}

\begin{document}

\title{Bridging Machine Learning and Glassy Dynamics Theory for Predictive Polymer Modeling}
\author{Anh D. Phan}
\email{anh.phanduc@phenikaa-uni.edu.vn}
\affiliation{Faculty of Materials Science and Engineering, Phenikaa University, Hanoi 12116, Vietnam}
\affiliation{Phenikaa Institute for Advanced Study, Phenikaa University, Hanoi 12116, Vietnam}
\author{Ngo T. Que}
\affiliation{Phenikaa Institute for Advanced Study, Phenikaa University, Hanoi 12116, Vietnam}
\author{Nguyen T. T. Duyen}
\affiliation{Faculty of Materials Science and Engineering, Phenikaa University, Hanoi 12116, Vietnam}
\author{Phan Thanh Viet}
\affiliation{Division of Physics, School of Education, Dong Thap University, Cao Lanh 870000, Vietnam}
\author{Quach K. Quang}
\affiliation{Division of Physics, School of Education, Dong Thap University, Cao Lanh 870000, Vietnam}
\author{Baicheng Mei}
\affiliation{School of Materials Science and Engineering, Beijing Institute of Technology, Beijing, 100081 China}
\date{\today}

\begin{abstract}
Understanding and predicting the glassy dynamics of polymers remain fundamental challenges in soft matter physics. While the Elastically Collective Nonlinear Langevin Equation (ECNLE) theory has been successful in describing relaxation dynamics, its practical application to polymers depends on a thermal mapping to connect theory with experiment, which in turn requires detailed thermodynamic data. Such data may not be available for chemically complex or newly designed polymers. In this work, we propose a simple approach that integrates machine learning-predicted glass transition temperatures ($T_g$) with a simplified thermal mapping based on an effective thermal expansion coefficient to overcome these limitations. This approach can provide quantitatively accurate predictions of relaxation dynamics across a broad range of polymers. Rather than replacing the original thermal mapping, our method complements it by trading formal rigor for computational efficiency and broader applicability in high-throughput screening and materials with limited available data. Moreover, we introduce a physically motivated modification to the thermal mapping that resolves discrepancies in the description of low-$T_g$ polymers. Our results establish a generalizable approach for predictive modeling of glassy polymer dynamics and point toward new directions for theory-guided materials discovery.
\end{abstract}

\keywords{Suggested keywords}
\maketitle
\section{Introduction}
Investigating the glassy dynamics of polymers, thermal liquids, organic materials, and amorphous drugs is critically important due to their broad applications across industrial, pharmaceutical, and technological sectors. Glassy dynamics strongly influence material stability, mechanical properties, and functional performance, particularly for amorphous pharmaceuticals where the glass transition temperature directly affects drug solubility, dissolution rates, and bioavailability \cite{1,2,3}. In polymers and organic materials, understanding structural relaxation processes and glass transition phenomena enables better control over processing conditions, mechanical durability, and thermal stability, essential for optimizing material performance in applications ranging from packaging to optoelectronics \cite{4,5}. Despite extensive experimental and theoretical studies, many fundamental aspects including the microscopic mechanisms governing alpha and beta relaxation processes remain incompletely understood, necessitating continued research to advance theoretical frameworks and predictive models \cite{6,7}.

To address these unresolved issues comprehensively, researchers have employed a variety of experimental and theoretical methods to investigate the activated dynamics of both supercooled liquids and glasses and the glass transition temperature. Experimentally, techniques such as differential scanning calorimetry (DSC), broadband dielectric spectroscopy (BDS), neutron scattering, and nuclear magnetic resonance have provided critical insights into relaxation behaviors, molecular mobility, and dynamic heterogeneities near the glass transition \cite{2,8}. Complementing these experiments, theoretical and computational approaches such as mode coupling theory, random first-order transition theory, and the ECNLE theory have provided powerful frameworks for interpreting and predicting the underlying physics of glass transition and structural relaxation dynamics \cite{7,9,10}. Recent advances in molecular dynamics (MD) simulations \cite{7,11,12} and machine/deep learning \cite{ML0, ML1, ML2, ML3, ML4, ML5, ML6, ML7, ML8, ML9, ML10} have further enhanced our ability to unravel microscopic mechanisms and quantitatively link experimental observations with theoretical predictions. These approaches significantly deepen and accelerate our understanding of glassy systems.

Among the theoretical frameworks, the ECNLE theory has emerged as a powerful and versatile approach for describing glassy dynamics across a wide range of materials \cite{7,11,12,13,14,15,16,17,18,19,20,21,22,23,24,25,26,27,28, polymer1, polymer2, polymer3, polymer4}. By quantitatively capturing the interplay between local cage-scale dynamics and longer-range collective elastic effects, the ECNLE theory has successfully explained the temperature dependence of structural relaxation times, viscosity growth, diffusion constants, and dynamic shear moduli in thermal liquids \cite{7, 12, 17, 18, 19, 21}, polymers \cite{11, 12, 13, 14, 15, 16, 20, 21, 22, 23, 27, polymer1, polymer2, polymer3, polymer4}, colloidal suspensions \cite{16, 17, 21}, metallic glasses \cite{25, 26, 28}, amorphous drugs \cite{24}, and nanocomposites \cite{25,26}. Furthermore, the ECNLE theory has been extended to investigate confinement effects in thin polymer films \cite{11,12,13,14,15,16,27}. The calculation of temperature-dependent relaxation times and diffusion constants within the ECNLE theory requires a thermal mapping from material density to temperature. This conversion is constructed from the experimental equation-of-state data or using experimental glass transition temperature values \cite{7,11,12,13,14,15,16,17,18,19,20,21,22,23,24,25,26,27}. Recent developments coupling ECNLE calculations with simulation data and machine learning models have further enhanced its predictive accuracy for metallic glasses \cite{26,28} without the need for adjustable or empirically fitted parameters.

Despite the significant successes of the ECNLE theory \cite{7,11,12,13,14,15,16,17,18,19,20,21,22,23,24,25,26,27} in describing the glassy dynamics of various materials, several critical challenges still hinder its broad predictive applicability. A major limitation lies in the dependence of ECNLE calculations on accurate knowledge of the glass transition temperature of materials, which are often unavailable for new or unsynthesized polymers. Traditional experimental determination of $T_g$ is time-consuming, costly, and infeasible for high-throughput material discovery. Furthermore, even when accurate $T_g$ values are available, the current ECNLE formulation struggles to accurately predict the structural relaxation behavior of certain polymers such as polydimethylsiloxane. This strongly suggests a need for further theoretical development concerning polymer-specific physics or mechanisms.

In this work, we systematically address the aforementioned challenges by combining machine learning (ML) techniques with the ECNLE theoretical framework. First, we train and validate several ML models on a comprehensive and chemically diverse dataset to accurately predict $T_g$. We then benchmark the predictive performance of these models against previous studies to identify the most effective ML approach for $T_g$ prediction. Subsequently, we integrate these ML-derived $T_g$ values into the ECNLE theory to quantitatively predict the temperature-dependent structural relaxation dynamics of a broad class of polymers. Finally, we propose a physically-motivated thermal mapping to accurately describe the temperature dependence of relaxation dynamics of polymers that were previously not well-captured. This work builds on our recent ML–ECNLE approach developed for metallic glasses and extends it to a wide range of polymer systems. Thereby, we significantly expand the predictive scope and generality of the ECNLE theory.

\section{Theoretical background}
\subsection{The ECNLE theory}
Within the framework of the ECNLE theory, the activation events in any amorphous material are theoretically investigated using a hard-sphere fluid model \cite{7,11,12,13,14,15,16,17,18,19,20,21,22,23,24,25,26,27,28}. Amorphous materials are modeled as a dense assembly of hard spheres characterized by the density of particle numbers, $\rho$, and the diameter of the particles, $d$. The static structure factor, $S(q)$, and the radial distribution function, $g(r)$, are calculated using the Percus–Yevick closure approximation \cite{29}. The local dynamics of a tagged particle is primarily governed by interactions with neighboring particles, random thermal noise, and a frictional force, as described by a nonlinear stochastic equation of motion. Solving this equation yields the dynamic free energy, $F_{\text{dyn}}(r)$, which quantifies the effective caging constraints experienced by the particle \cite{7,11,12,13,14,15,16,17,18,19,20,21,22,23,24,25,26,27,28}

\begin{eqnarray} 
\frac{F_{\text{dyn}}(r)}{k_B T} &=& -3\ln \left(\frac{r}{d}\right) \\
&-& \int_0^\infty \frac{q^2 d^3 [S(q)-1]^2}{12\pi \Phi [1 + S(q)]} \exp\left[-\frac{q^2 r^2 (1 + S(q))}{6 S(q)}\right] dq,\nonumber
\label{eq:1}
\end{eqnarray}
where $T$ is the temperature, $k_B$ is the Boltzmann constant, $r$ is the particle displacement, $q$ is the wavevector, and $\Phi = \rho \pi d^3 / 6$ is the volume fraction. The first term on the right-hand side of Eq.~(\ref{eq:1}) describes ideal fluid-like behavior, while the second term accounts for the confinement effect due to neighboring particles (caging constraint).

\begin{figure*}[htp]
\includegraphics[width=16cm]{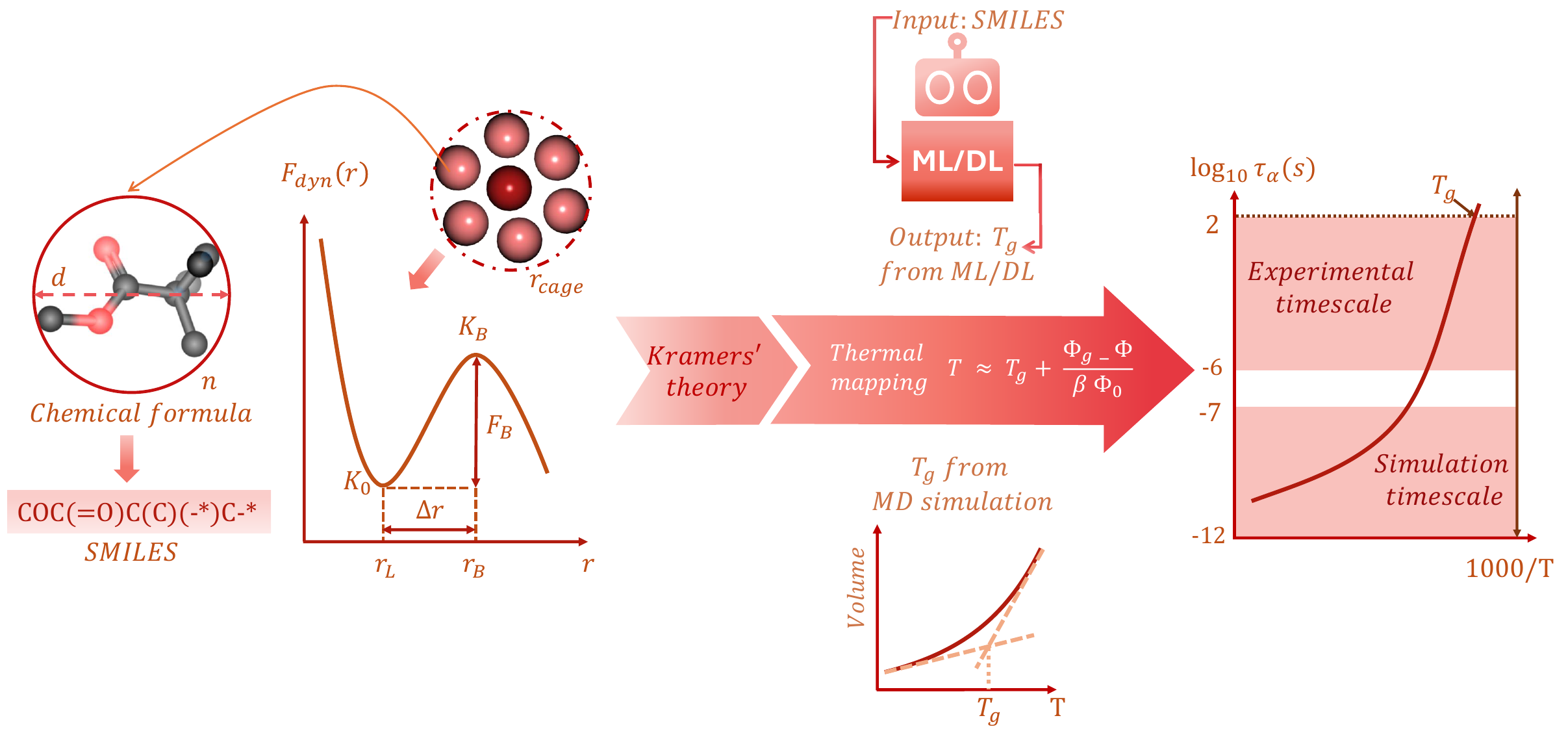}
\caption{\label{fig1}(Color online) Workflow for predicting the glass transition temperature of polymers using molecular dynamics simulations and deep learning/machine learning, after which the predicted $T_g$ is used in the ECNLE theory to compute the structural relaxation time and diffusion coefficients.}
\end{figure*}

The dynamic free energy provides critical understanding of the mechanisms that govern the local motion of the particles. At low volume fractions ($\Phi < 0.432$), the particles exhibit the characteristic behavior of an ideal fluid and freely diffuse with minimal resistance due to weak interactions between particles \cite{15,16,17,18,19}. As the system becomes denser, stronger interparticle forces lead to the emergence of cage-like structures that transiently trap a central particle. The characteristic size of this cage, $r_{\text{cage}}$, is identified as the position of the first minimum in the radial distribution function, $g(r)$. Consequently, the local energy barrier in the dynamic free energy landscape increases as illustrated in Fig.~\ref{fig1}. The height of this barrier is determined as $F_B = F_{\text{dyn}}(r_B) - F_{\text{dyn}}(r_L)$, where $r_L$ and $r_B$ correspond to the localization length and the barrier position of $F_{\text{dyn}}(r)$, respectively. The displacement between these two dynamic positions, $\Delta r = r_B - r_L$, characterizes the jump distance of the particle required to escape the local cage.

To successfully escape its local cage, a particle must overcome not only its immediate constraints but also facilitate cooperative rearrangements of the surrounding particles to generate sufficient free volume \cite{7,11,12,13,14,15,16,17,18,19,20,21,22,23,24,25,26,27,28}. This collective motion produces a radially outward displacement field, $u(r)$, which originates at the cage surface and propagates into the surrounding medium. Within the framework of Lifshitz’s continuum elasticity theory \cite{30}, the displacement field for $r \geq r_{\text{cage}}$ can be analytically expressed as

\begin{eqnarray} 
u(r) = \frac{\Delta r_{\text{eff}} r_{\text{cage}}^2}{r^2}, 
\label{eq:2}
\end{eqnarray}
where $\Delta r_{\text{eff}}$ is the effective amplitude of the displacement field. The analytical expression of $\Delta r_{\text{eff}}$ is \cite{17,18,19}
\begin{eqnarray} 
\Delta r_{eff} = \frac{3}{r_{cage}^3}\left[\frac{r_{cage}^2\Delta r^2}{32} - \frac{r_{cage}\Delta r^3}{192} + \frac{\Delta r^4}{3072} \right].
\label{eq:2-1}
\end{eqnarray}

Since $u(r)$ is generally small, the particle displacements are treated as harmonic oscillations. The local elastic energy for a single oscillator is $K_0 u^2(r)/2$, where $K_0 = \left|\partial^2 F_{dyn}(r)/\partial r^2\right|_{r=r_L}$ is a  packing-fraction-dependent spring constant. The total collective elastic barrier is obtained by integrating the elastic energy contributions from all particles located outside the cage \cite{7,11,12,13,14,15,16,17,18,19,20,21,22,23,24,25,26,27,28}
\begin{eqnarray} 
F_e = 4\pi \rho \int_{r_{\text{cage}}}^{\infty} r^2 g(r)\frac{K_0u^2(r)}{2}dr. 
\label{eq:3}
\end{eqnarray}

The overall structural relaxation time, $\tau_\alpha$, is then computed using Kramers’ theory by incorporating both local and collective barrier contributions as \cite{7,11,12,13,14,15,16,17,18,19,20,21,22,23,24,25,26,27,28}
\begin{eqnarray} 
\frac{\tau_\alpha}{\tau_s} = 1 + \frac{2\pi}{\sqrt{K_0 K_B}} \frac{k_B T}{d^2} \exp\left(\frac{F_B + F_e}{k_B T}\right), 
\label{eq:4}
\end{eqnarray}
where $\tau_s$ is a short-timescale prefactor related to the combined effects of binary collision rate and weak many body caging and $K_B$=$\left|\left.\partial^2 F_{dyn}(r)/\partial r^2\right|_{r=r_B}\right|$ is the absolute curvature at the barrier position. The explicit form of $\tau_s$ was provided in previous studies \cite{15,16,17,18}.

Equation (\ref{eq:4}) yields $\tau_\alpha$ as a function of volume fraction. However, direct comparison with experimental data requires mapping density to temperature. Originally, Mirigian and Schweizer proposed a thermal mapping framework for rigid molecular liquids \cite{18,20}, which uses a more thermodynamic basis. This approach is based on reproducing the experimental long-wavelength dimensionless density fluctuation amplitude, $S_{0}^{\text{expt}}(T) = \rho k_B T \kappa_T$, where $\kappa_T$ is the isothermal compressibility. This quantity captures nanoscale density fluctuations. This quantity is a sensitive probe of nanoscale density fluctuations and is derived from the experimental equation of state (EOS). Within the hard-sphere model using the Percus-Yevick closure, $S_0$ is given analytically as $\cfrac{(1 - \Phi)^4}{(1 + 2\Phi)^2}$. By equating this expression with an empirical fit to EOS data \cite{13,14,15,16,17,18,19,20,21,22,23}
\begin{eqnarray}
S_0^{HS} = \frac{(1 - \Phi)^4}{(1 + 2\Phi)^2} \equiv S_{0}^{\text{expt}}(T) \approx \frac{1}{N_s}\left(-A + \frac{B}{T}\right)^{-2},
\label{eq:6}
\end{eqnarray}
where $A$ and $B$ are empirically derived EOS parameters representing entropy and cohesive energy contributions, respectively, and $N_s$ is the number of interaction sites per molecule such as chemical groups of -C-, -CH-, -O-, -NH-, and so on. From this, one obtains the effective hard-sphere packing fraction as a function of temperature as \cite{20,21}
\begin{eqnarray}
\Phi_{\text{eff}}(T) = 1 + \sqrt{S_{0}^{expt}(T)} - \sqrt{S_{0}^{expt}(T) + 3 \sqrt{S_{0}^{expt}(T)}}.
\label{eq:7}
\end{eqnarray}
This approach requires knowledge of molecular-specific parameters ($A$, $B$, and $N_s$). Once $\Phi_{\text{eff}}(T)$ is obtained, we can calculate $g(r)$ and $S(q)$ and hence $F_{\text{dyn}}(r)$ and the temperature dependence of the structural relaxation time $\tau_\alpha(T)$ without any adjustable parameters. 

While the original ECNLE theory has successfully predicted glassy dynamics across more than 14 decades in relaxation time for diverse molecular and polymeric glass-formers \cite{13,14,15,16,17,18,19,20,21,22,23}, it assumes a universal coupling between local cage dynamics and long-range elasticity. This assumption limits its ability to describe the significant variability in dynamic fragility observed among chemically distinct polymers. To overcome this limitation, Xie and Schweizer introduced a modified ECNLE approach incorporating a non-universal coupling between local cage-scale hopping and the longer-range collective elastic distortions \cite{31}. Specifically, they assumed that the microscopic jump distance required for a local cage escape should be chemically specific and influenced by nanometer-scale conformational transitions and monomer-level interactions. This chemical specificity was implemented through a single material-dependent parameter $a_c$which scales the collective elastic barrier in Eq. (\ref{eq:3}) as $F_{e,new} \equiv a_c F_e$. From this, Eq. (\ref{eq:4}) becomes
\begin{eqnarray}
\frac{\tau_\alpha}{\tau_s} = 1 + \frac{2\pi}{\sqrt{K_0 K_B}} \frac{k_B T}{d^2} \exp\left(\frac{F_B + a_cF_e}{k_B T}\right). 
\label{eq:8}
\end{eqnarray}
The extended theory significantly improves simultaneous predictions for the glass transition temperature, the dynamic fragility, and the temperature dependence of $\tau_\alpha$.

Despite these improvements, practical application of the extended theory is limited by the challenge of accurately determining EOS parameters and $N_s$ for long-chain or chemically complex polymers. To address this limitation, we proposed an alternative thermal mapping in our prior works \cite{24,25,26,28}, based on the principle of thermal expansion, which is

\begin{eqnarray} 
T \approx T_g + \frac{\Phi_g - \Phi}{\beta \Phi_0},
\label{eq:5}
\end{eqnarray}
where $\beta \approx 12\times10^{-4} \text{K}^{-1}$ is an effective thermal expansion coefficient (TEC), $\Phi_0 = 0.5$ is a characteristic packing fraction, and $\Phi_g$ corresponds to the volume fraction at the glass transition state defined when $\tau_\alpha(\Phi_g) = \tau_\alpha(T_g)= 100$ s. Obviously, $\Phi_g$ depends on the magnitude of the material-specific parameter $a_c$, which scales the collective elastic barrier. In the case of $a_c = 1$, the glass transition packing fraction is set to $\Phi_g = 0.6157$, consistent with previous studies \cite{24,25,26,28}. Our simplified thermal mapping in Eq.~(\ref{eq:5}) provides a simpler and more practical alternative to the original mapping approach \cite{13,14,15,16,17,18,19,20,21,22,23,31}. The original mapping approach is physically rigorous, and a recent work \cite{polymer5} has linked key thermodynamic parameters to the fragility and glass transition temperature of molecular and polymer liquids. However, this method requires detailed input such as $A$, $B$, and $N_s$, which are often difficult to determine experimentally, especially for complex or novel polymer systems \cite{24,25,26,28,polymer5}. In contrast, our method is easier to apply to large datasets or systems lacking full thermodynamic characterization. Despite its simplicity, this mapping still reflects the essential relationship between $T_g$ and thermal expansion as reported in prior studies \cite{51,34}. This agreement reinforces the view that thermal expansivity encodes essential molecular-scale information governing the glass formation. The $T_g$ value can be determined through DSC and/or BDS experiment, or alternatively estimated via computational simulations or machine learning \cite{ML0, ML1, ML2, ML3, ML4, ML5, ML6, ML7, ML8, ML9, 26, 28,50} as indicated in Fig. \ref{fig1}. 

\subsection{Machine learning calculations}
To implement the thermal mapping described in Eq.(\ref{eq:5}) to investigate a new polymer, machine learning models can be exploited to predict the glass transition temperature. In this work, we utilize a large-scale dataset compiled in Ref. \cite{ML5}, which contains 7174 polymers with experimentally measured $T_g$ values and their corresponding chemical structures encoded using the Simplified Molecular Input Line Entry System (SMILES) notation. The dataset spans a wide range of $T_g$ values from 134 K to 768 K with an average of approximately 415.1 K. To uncover structure–property relationships and predict $T_g$ values for new polymers, we apply four widely used ML algorithms including Gaussian Process Regression (GPR), Extra Trees Regression (ETR), Extreme Gradient Boosting (XGB), and Random Forest Regression (RFR). These models were selected for their proven ability to handle high-dimensional, non-linear relationships commonly observed in polymer datasets. Tree-based ensemble methods provide interpretability through feature importance measures \cite{28,ML5}, while GPR is well-suited for capturing complex trends with limited data through its probabilistic framework \cite{ML0,32}. For the GPR model, we follow the implementation procedure outlined in Ref.\cite{32}. Meanwhile, for tree-based models, we adopted the methodology detailed in Ref. \cite{28}. 

To evaluate model performance and quantify the predictive relationship between polymer structure and $T_g$, we trained the models using features derived from the SMILES representations. The dataset was randomly split into training and testing subsets using an 80:20 ratio. The model accuracy was assessed using the coefficient of determination ($R^2$) and the root mean square error (RMSE) defined as
\begin{eqnarray} R^2 &=& 1 - \frac{\sum_{i=1}^{n}(y_i - \hat{y}i)^2}{\sum{i=1}^{n}(y_i - \bar{y})^2}, \\ 
\text{RMSE} &=& \sqrt{\frac{1}{n} \sum_{i=1}^{n}(y_i - \hat{y}_i)^2}, 
\end{eqnarray}
where $y_i$ and $\hat{y}_i$ are the observed and predicted $T_g$ values for the $i$-th polymer, $\bar{y}$ is the mean of the observed values, and $n$ is the number of data points. Higher $R^2$ values and lower RMSE values indicate better model performance and predictive accuracy.

\section{Results AND Discussion}
\begin{figure*}[htp]
\includegraphics[width=16cm]{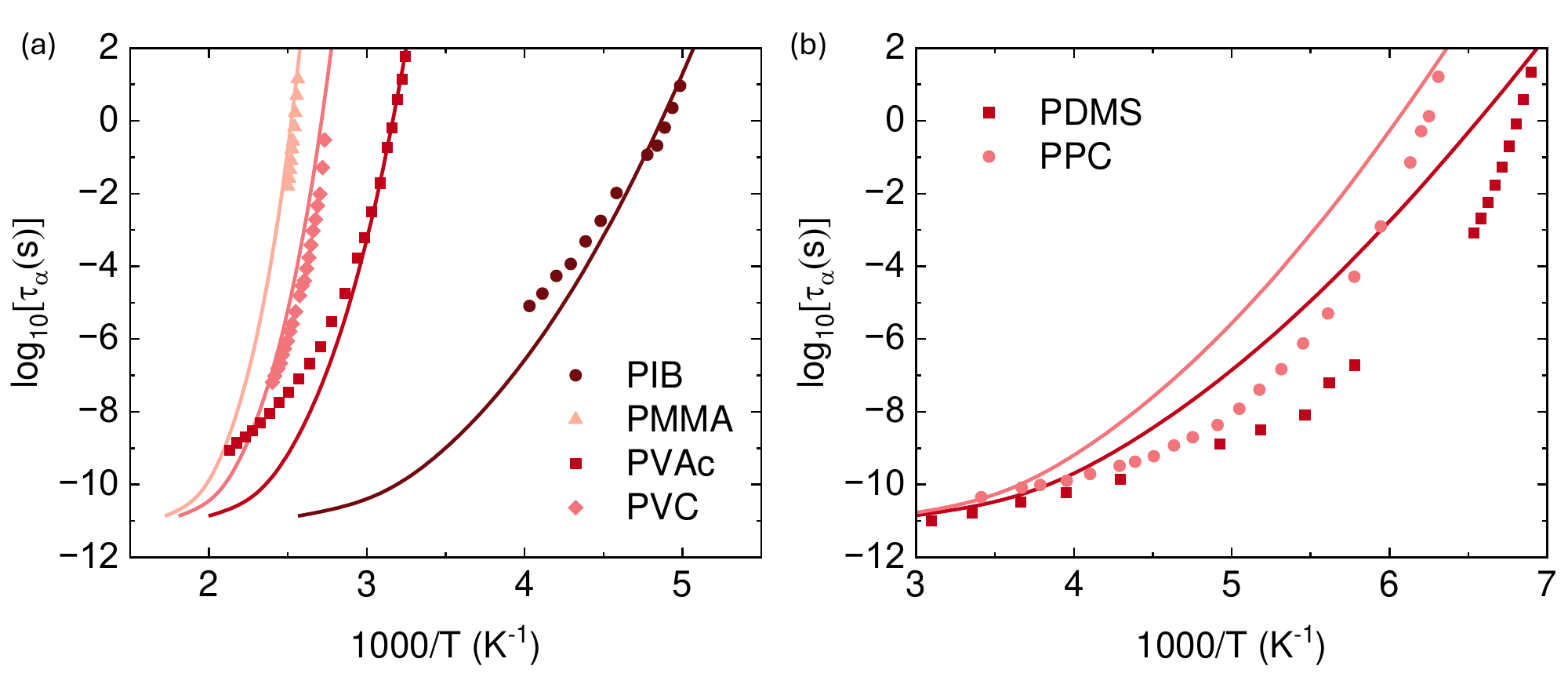}
\caption{\label{fig2}(Color online) Temperature dependence of logarithm of structural relaxation time of (a) polyisobutene (PIB), poly(methyl methacrylate) (PMMA), polyvinyl acetate (PVAc), and polyvinyl chloride (PVC), and (b) polydimethylsiloxane (PDMS) and polypropylene carbonate (PPC). Data points and solid curves correspond to BDS experimental data in prior works \cite{Tg_relaxation_PIB, Tg_relaxation_PMMA, Tg_relaxation_PVAc, Tg_relaxation_PVC, Tg1, Relaxation_PPC_H20:LiCl} and ECNLE calculations, respectively. The $T_g$ values are determined using the BDS data.}
\end{figure*}

Figure \ref{fig2} compares experimental structural relaxation times, $\tau_\alpha(T)$, and with predictions from the ECNLE theory (Eqs.(1-5) and (\ref{eq:5})) for several representative polymers. As shown in Fig. \ref{fig2}a, the theoretical calculations quantitatively reproduce the experimental data over a broad temperature range for PIB, PMMA, PVAc, and PVC. This is consistent with previous observations for many amorphous materials \cite{24,25,26,28}. This agreement validates the ability of the ECNLE framework to capture the essential features of glassy dynamics without any adjustable parameter. These polymers have also been previously studied  using the $S_0(T)$-based mapping with Kuhn segment-level coarse-graining in Refs.~\cite{20,31}, where the ECNLE theory was able to quantitatively describe fragility and relaxation over wide temperature ranges. Our simplified thermal expansion-based mapping provides a practical alternative in cases where detailed thermodynamic input is not available and still produces accurate results for these polymers. Our approach complements previous methods by trading some theoretical rigor for broader computational applicability. 

In contrast, substantial deviations between the theoretical predictions and experimental data are observed for PDMS and PPC as seen in Fig. \ref{fig2}b. Although accurate prediction of the glass transition temperature is essential to initiate the thermal mapping , $T_g$ alone does not fully dictate the temperature dependence of structural relaxation. These discrepancies likely arise from specific molecular characteristics including intramolecular flexibility, side-chain dynamics, and specific intermolecular interaction. To account for these effects, we adopt the approach proposed by Xie and Schweizer by introducing a material-specific parameter $a_c$ that adjusts the importance of the collective elastic contribution to the total barrier ($F_B+ a_c F_e$) \cite{31,24}.  Unlike thermodynamic approaches that rely on energy landscapes or packing structures, this modification directly captures how collective molecular motions slow down particle movement. The value of $a_c$ is treated as an adjustable parameter to obtain the best fit to the experimental data \cite{31,24}.

\begin{figure*}[htp]
\includegraphics[width=16cm]{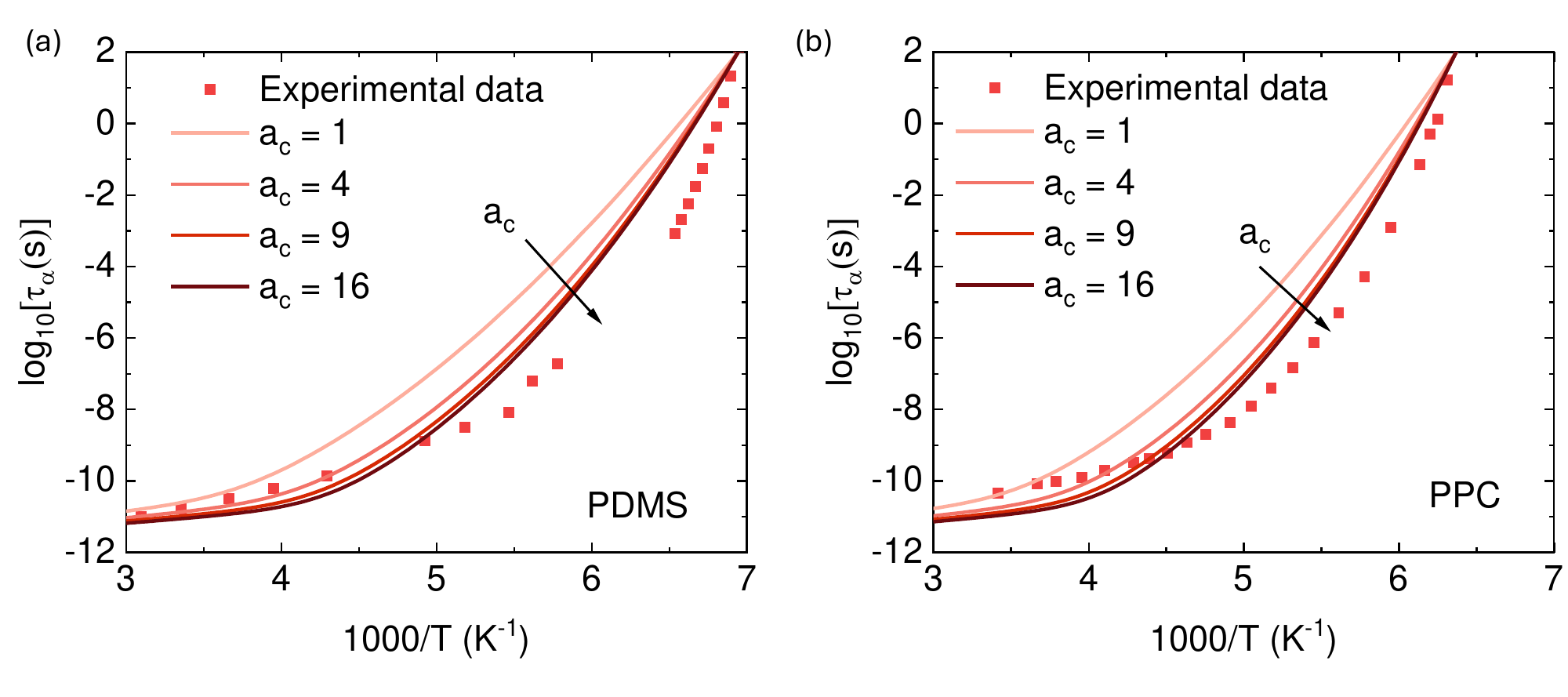}
\caption{\label{fig3}(Color online) Temperature dependence of logarithm of structural relaxation time of (a) PDMS and (b) PPC. Data points and solid curves correspond to BDS experimental data in prior works \cite{Tg1, Relaxation_PPC_H20:LiCl} and ECNLE calculations with various values of $a_c$, respectively. The $T_g$ values are determined using the BDS data.}
\end{figure*}

Figure \ref{fig3} shows ECNLE calculations (Eqs. (1-4), (\ref{eq:8}), and (\ref{eq:5})) for PDMS and PPC using different values of the $a_c$ parameter and compares the results with experimental data. Although tuning $a_c$ leads to better agreement, the predictions still fall short of quantitatively capturing the relaxation times across the full temperature range. Remarkably, the earlier ECNLE work assuming $a_c = 1$ \cite{20} when using the more detailed $S_0(T)$-based thermal mapping and Kuhn segment coarse-graining, showed that despite underestimating the $T_g$ value of PDMS, the model successfully captured key features such as the molecular weight dependence of $T_g$ (normalized by its infinite molecular weight limit, $T_g(\infty)$) and the temperature dependence of $\tau_\alpha$ in an Angell plot. In contrast, when our results for PDMS and PPC in Fig. \ref{fig3} are replotted in the Angell representation (see Fig. S1 in the Supplementary Materials), the theoretical curves do not quantitatively agree experimental data. This comparison suggests that the ECNLE calculations based on the simplified thermal expansion mapping [Eq.~(\ref{eq:5})] are not sufficient to fully capture the relaxation dynamics of polymers with complex molecular structures. While an accurate estimate of $T_g$ is important, it does not guarantee a reliable prediction of the temperature-dependent behavior of structural relaxation. Further theoretical development is needed. 

\begin{figure*}[htp]
\includegraphics[width=16cm]{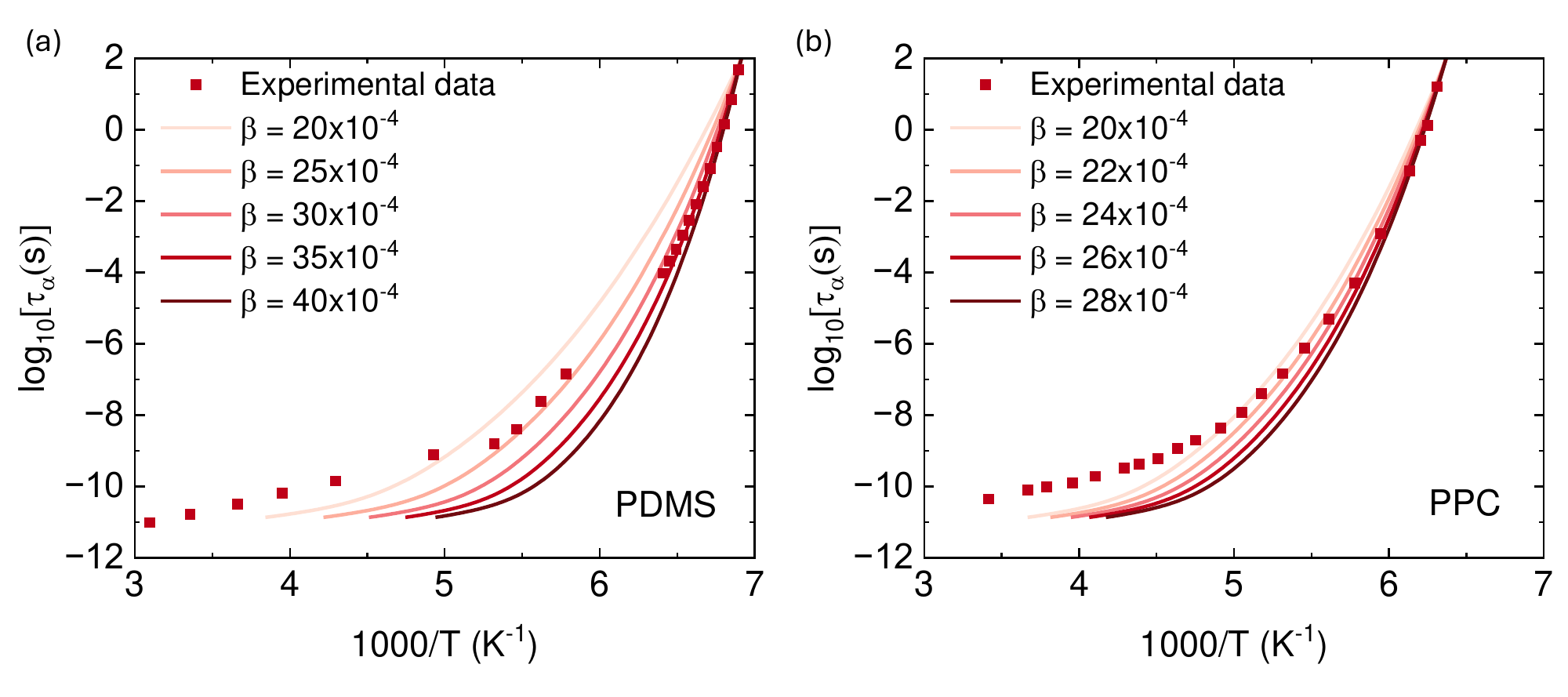}
\caption{\label{fig4}(Color online)  Temperature dependence of logarithm of structural relaxation time of (a) PDMS and (b) PPC. Data points and solid curves correspond to BDS experimental data in prior works \cite{Tg1, Relaxation_PPC_H20:LiCl} and ECNLE calculations with various values of $\beta$ in Eq. (\ref{eq:5}), respectively. The $T_g$ values are determined using the BDS data.}
\end{figure*}

In recent studies \cite{53,54}, Schweizer and coworkers employed the Polymer Reference Interaction Site Model (PRISM) theory to calculate the static structure factor and direct correlation function for discrete semiflexible polymer chains. This approach explicitly accounts for chain connectivity and conformational stiffness, which are important features of polymer dynamics. The resulting structural information is then used to construct the dynamic free energy and compute relaxation times following the same procedures as in the ECNLE theory based on the hard-sphere fluid model discussed in Section II. As shown in Fig. S1 of the Supplementary Materials, the predicted temperature dependence of the structural relaxation time from both the hard-sphere and polymer-specific models shows excellent agreement, even for varying chain lengths. This result suggests that the hard-sphere model, despite its simplicity, effectively captures the essential physics of glassy relaxation in polymer melts and remains a robust and practical choice for providing the alternative method to the $S_0$ mapping idea, which both can be combined with the ENCLE theory, when detailed structural data are not available.

An important assumption in our simplified thermal mapping (Eq.(\ref{eq:5})) is that the TEC of polymers remains nearly constant. This assumption is well-supported by experimental data reported in Ref.\cite{34}, where polymers with $T_g > 160$ K have remarkably similar thermal expansion behavior. Such consistency suggests that, above this $T_g$ threshold, the volumetric response of polymers is governed by general thermodynamic constraints and is relatively insensitive to specific chemical structure. However, for polymers with lower glass transition temperatures ($T_g < 160$ K), including PDMS and PPC, the situation appears to deviate from this trend. These materials exhibit significantly greater chain flexibility and lower cohesive energy densities, which are expected to result in enhanced thermal expansivity. As a result, although the TEC in this low-$T_g$ regime is still treated as constant to a zeroth-order approximation, it is hypothesized to be greater than that of high-$T_g$ polymers. This hypothesis provides a plausible explanation for the discrepancies observed in ECNLE predictions for PDMS and PPC when using the standard thermal mapping parameters.

To test this hypothesis and assess its impact on ECNLE predictions, we perform new calculations for PDMS and PPC. Using Eq.(\ref{eq:4}) to compute $\tau_\alpha(\Phi)$, we re-apply the thermal mapping in Eq.(\ref{eq:5}) with an increased value of  $\beta$ to mimic the anticipated higher TEC in these systems. As shown in Fig. \ref{fig4}, this adjustment yields ECNLE predictions of $\tau_\alpha(T)$ that are in excellent quantitative agreement with experimental BDS data. The predictive capability of the ECNLE theory is substantially enhanced relative to the case employing a universal $\beta$ value. These results strongly support the notion that low-$T_g$ polymers have significantly greater thermal expansion and validate the need for an adjusted mapping in such systems. However, due to the lack of direct experimental measurements of TEC in this regime, this remains a working hypothesis. Future experimental investigations aimed at accurately quantifying the thermal expansion behavior of low-$T_g$ polymers would be critical to confirm this assumption. To partially assess the plausibility of our chosen value, in Ref.~\cite{31}, $A = 0.675$, $B=1057$, and $N_s=12.6$ for PDMS, we calculate $\Phi_{eff}(T)$ based on Eq. (\ref{eq:6}) and find that the slope of $\Phi_{eff}(T)$ is approximately $7.5 \times 10^{-4}$ K$^{-1}$ at high temperatures and around $10 \times 10^{-4}$ K$^{-1}$ at lower temperatures. It corresponds to the effective thermal expansion coefficient ($\beta$ in our thermal mapping) in the range of $15$–$20 \times 10^{-4}$ K$^{-1}$, which is reasonably close to our employed value. Consequently, as a zeroth-order approximation, we adopt a thermal expansion coefficient of $\beta = 24 \times 10^{-4}$ K$^{-1}$, which is twice the universal value previously applied to all materials \cite{7,11,12,13,14,15,16,17,18,19,20,21,22,23,24,25,26,27,28}, for polymers with predicted $T_g$ values below 160 K. This comparison provides further justification that our use of a higher $\beta$ value for low-$T_g$ polymers is not arbitrary, but rather grounded in available experimental data.

\begin{figure*}[htp]
\includegraphics[width=16cm]{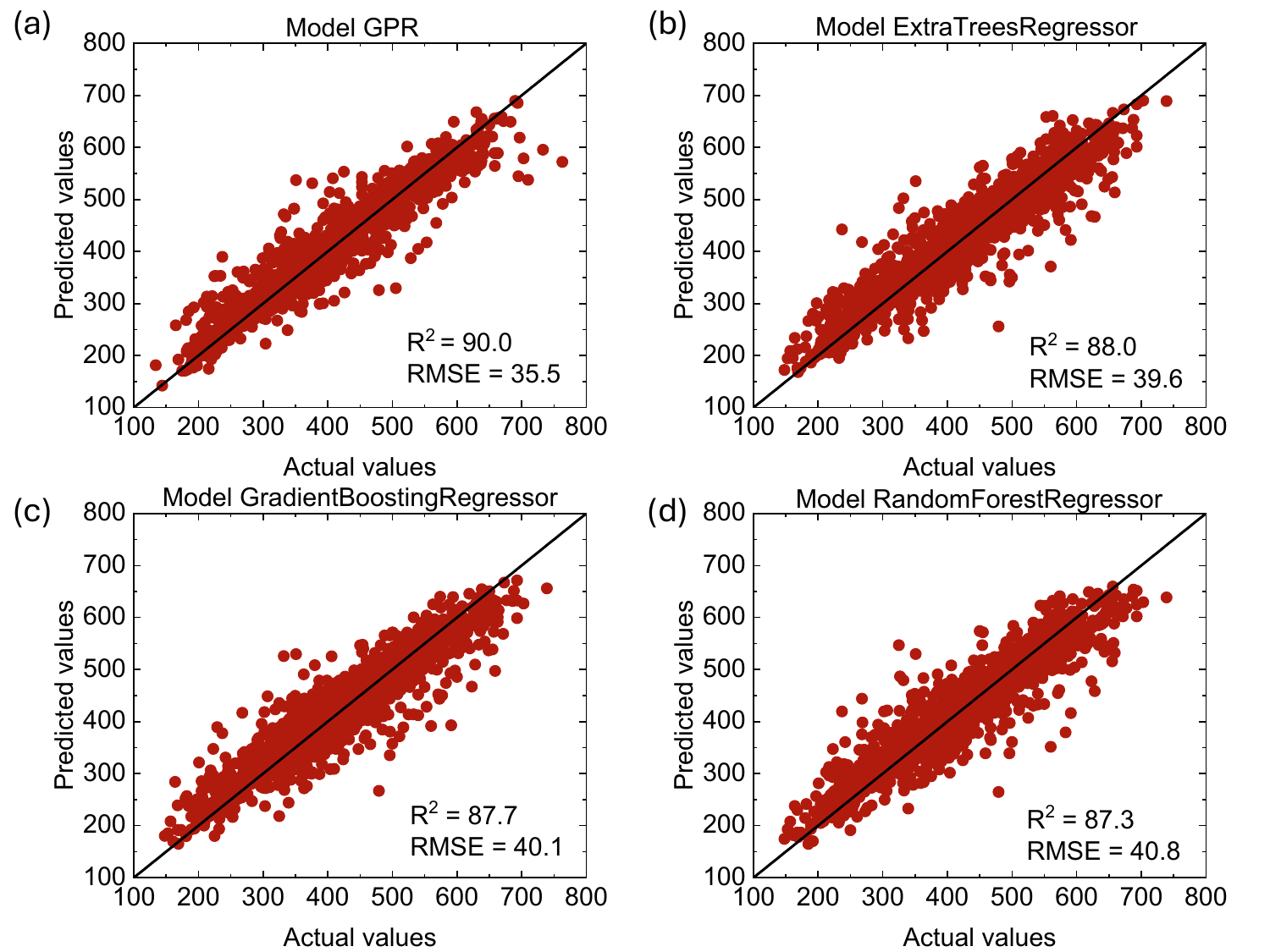}
\caption{\label{fig5}(Color online) Predictive performance of four regression models on the testing dataset for polymer $T_g$ prediction: (a) Gaussian Process Regression, (b) Extra Trees Regression, (c) Extreme Gradient Boosting, and (d) Random Forest Regression.}
\end{figure*}

The results presented above indicate that ECNLE calculations associated with the experimental $T_g$ values yield excellent quantitative agreement with BDS data. This raises an important question: can comparable predictive accuracy be achieved when substituting experimental $T_g$ values with those obtained from machine learning models? To address this, we trained four regression models (GPR, ETR, XGB, and RFR) on the polymer dataset reported in Ref.\cite{ML5}. As shown in Fig.\ref{fig5}, GPR achieved the highest predictive performance with an $R^2$ value of 90$\%$ and an RMSE of approximately 35.5 K. This outperforms previously reported models applied to the same dataset (Table~\ref{tab:ml_dl_models}), where the best $R^2$ was 86.9$\%$ and most RMSE values exceeded 40 K. Although ETR, XGB, and RFR in our study also surpassed their counterparts in previous works \cite{ML5}, their $R^2$ values remained below 89$\%$. Increasing the training set generally leads to modest improvements in model accuracy, as shown in Table SI (supplemental materials). Among all models tested, GPR consistently exhibits the best predictive performance in different train-test splits with $R^2 > 89$ $\%$ and RMSE below 37 K.

To further assess the accuracy of our machine learning–assisted ECNLE predictions of glass transition temperature, we compare them with results from previous ECNLE studies that employed the $S_0(T)$-based thermal mapping as discussed in Refs.~\cite{20,31}. In Ref. \cite{20}, Schweizer and coworkers employed the ECNLE theory in conjunction with the $S_0(T)$-based thermal mapping (Eq. (\ref{eq:7})) and assumed $a_c =1$ to predict $T_g$ values for 12 polymers. Their approach yielded RMSE of 46.32 K and $R^2$ of 76.34 $\%$ against experimental $T_g$ values. In a later study \cite{31}, they introduced a specific chemistry parameter $a_c$ to improve the predictive accuracy of both $T_g$ and the dynamic fragility of 17 polymers and achieve RMSE $\approx$ 33.1 K and $R^2\approx 86\%$. In comparison, our machine learning approach obtains similar or better predictive accuracy without the need for adjustable parameters and is trained on a much larger and more chemically diverse polymer dataset. Although the ECNLE theory provides important physical insights and captures the dependence of molecular weight on $T_g$, our data-driven approach provides an alternative to link the ECNLE theory to real materials, especially for high-throughput applications or systems that lack detailed thermodynamic input.

Note that certain specialized approaches in prior studies achieved even higher predictive accuracy but they have inherent limitations. Zhan and his coworkers \cite{ML4} attained an exceptional predictive accuracy ($R^2 \approx 97\%$) using the XGB algorithm combined with data augmentation techniques. However, their dataset was constructed exclusively from simulation data which lack the complexity and variability observed in experimental measurements. Consequently, their approach may not yield comparable accuracy when applied to experimental data. Similarly, Dong et al.\cite{ML3} integrated graph convolutional networks and reinforcement learning to achieve a predictive accuracy of nearly $93\%$. However, this hybrid method is considerably more complex and their dataset primarily consisted of structurally similar polyimides. Clearly, the dataset lacks the chemical diversity present in our broader dataset. As a result, their high predictive accuracy might diminish if applied to a chemically diverse polymer dataset. These comparisons emphasize the strength of our GPR model in reliably capturing complex, nonlinear structure–property relationships across chemically diverse, experimentally derived polymer datasets due to its probabilistic formulation and flexibility in modeling smooth functions.

\begin{table}[htbp]
\centering
\caption{Summary of datasets, ML and DL models, and their corresponding $R^2$ and RMSE for the glass transition temperature prediction in previous works. The ML/DL models include Gaussian Process Regression (GPR), Graph Convolutional Network (GCN), Convolutional Neural Network (CNN), Generative Adversarial Networks (GANs), Molecular Structural Regularized Graph Convolutional Network with Reinforcement Learning (MSRGCN-RL), Extreme Gradient Boosting (XGB), Ridge Regression (RR), Least Absolute Shrinkage and Selection Operator (LASSO), Support Vector Regression (SVR), Decision Tree  (DT), AdaBoost (AB), K-Nearest Neighbors (KNN), Random Forest Regression (RFR), Light Gradient Boosting Machine (LGB), Histogram-Based Gradient Boosting (HGB), Extra Tree Regression (ETR), Artificial Neural Network (ANN), Recurrent Neural Network (RNN), and Graph Convolutional Neural Network (GCNN).}
\begin{tabular}{|c|l|c|c|c|}
\hline
\textbf{Size of data} & \textbf{ML/DL model} & \textbf{R\textsuperscript{2}} & \textbf{RMSE} & \textbf{Ref} \\
\hline
5076  & GPR                           &  -    & 35.3     & \cite{ML0} \\\hline
300   & GCN                           & 88    & 19.4     & \cite{ML1} \\\hline
$>$400& CNN                           &  $<$ 80    & 49       & \multirow{2}{*}{\cite{ML2}} \\
      & GANs                          &  $<$ 92     & 31       &            \\\hline
1200  & MSRGCN-RL                     & 92.76 & 12.07    & \cite{ML3} \\\hline
800   & XGB                           & 97.10 & 0.736    & \multirow{4}{*}{\cite{ML4}} \\
      & RR                         & 84.49 & 2.542    &              \\
      & LASSO                         & 84.48 & 2.548    &              \\
      & SVR                           & 96.51 & 1.057    &              \\\hline
7174  & DT                            & 73.4  & 57.299   & \multirow{9}{*}{\cite{ML5}} \\
      & SVR                           & 86.1  & 41.36   &              \\
      & AB                            & 69.0  & 61.920   &              \\
      & KNN                           & 82.9 & 45.963   &              \\
      & XGB                           & 86.2  & 41.249   &              \\
      & RFR                            & 82.5  & 46.507   &              \\
      & LGB                           & 85.9  & 41.773   &              \\
      & HGB                           & 86.1  & 41.371   &              \\
      & ETR                           & 86.9  & 40.286   &              \\\hline
1437  & KNN                           & 71.0  &   -      & \multirow{5}{*}{\cite{ML6}} \\
      & SVR                           & 68.9  &   -      &              \\
      & XGB                           & 77.4  &   -      &              \\
      & ANN                           & 79.0  &   -      &              \\
      & RNN                           & 76.3  &   -      &              \\\hline
60    & GPR                           &  -     & 2.2762   & \cite{ML7} \\\hline
6923  & CNN                           & 83    & 48.45    & \multirow{3}{*}{\cite{ML8}} \\
      & RNN                           & 81    & 45.81    &              \\
      & GCNN                          & 88    & 38.82    &              \\\hline
902   & SVR                           & 77    &   -      & \cite{ML9} \\ \hline
$\sim$ 7500 & GCN                           & 90    &   38.1      & \cite{ML10} \\
\hline
\end{tabular}
\label{tab:ml_dl_models}
\end{table}

A recent study led by Glova and Karttunen \cite{50} presents a machine learning framework that utilizes dimensionality reduction of MD simulation data to estimate the glass transition temperature of polymers. Their method combines nonlinear diffusion maps, principal component analysis, and Gaussian mixture modeling to determine the $T_g$ value from the radial distribution functions, mean square displacements, relative square displacements, and dihedral angles. While this approach provides valuable physical insight into glassy dynamics from atomistic simulations, its reliance on system-specific and high-resolution MD trajectories restricts the applicability for high-throughput screening and the evaluation of unsynthesized polymers. In contrast, our study introduces a data-driven framework that predicts $T_g$ directly from the chemical structure of the polymer using experimental data and composition-based descriptors. Thus, this allows accurate and computationally efficient $T_g$ prediction and is well suited for material discovery and screening in broad chemical spaces.

\begin{figure*}[htp]
\includegraphics[width=16cm]{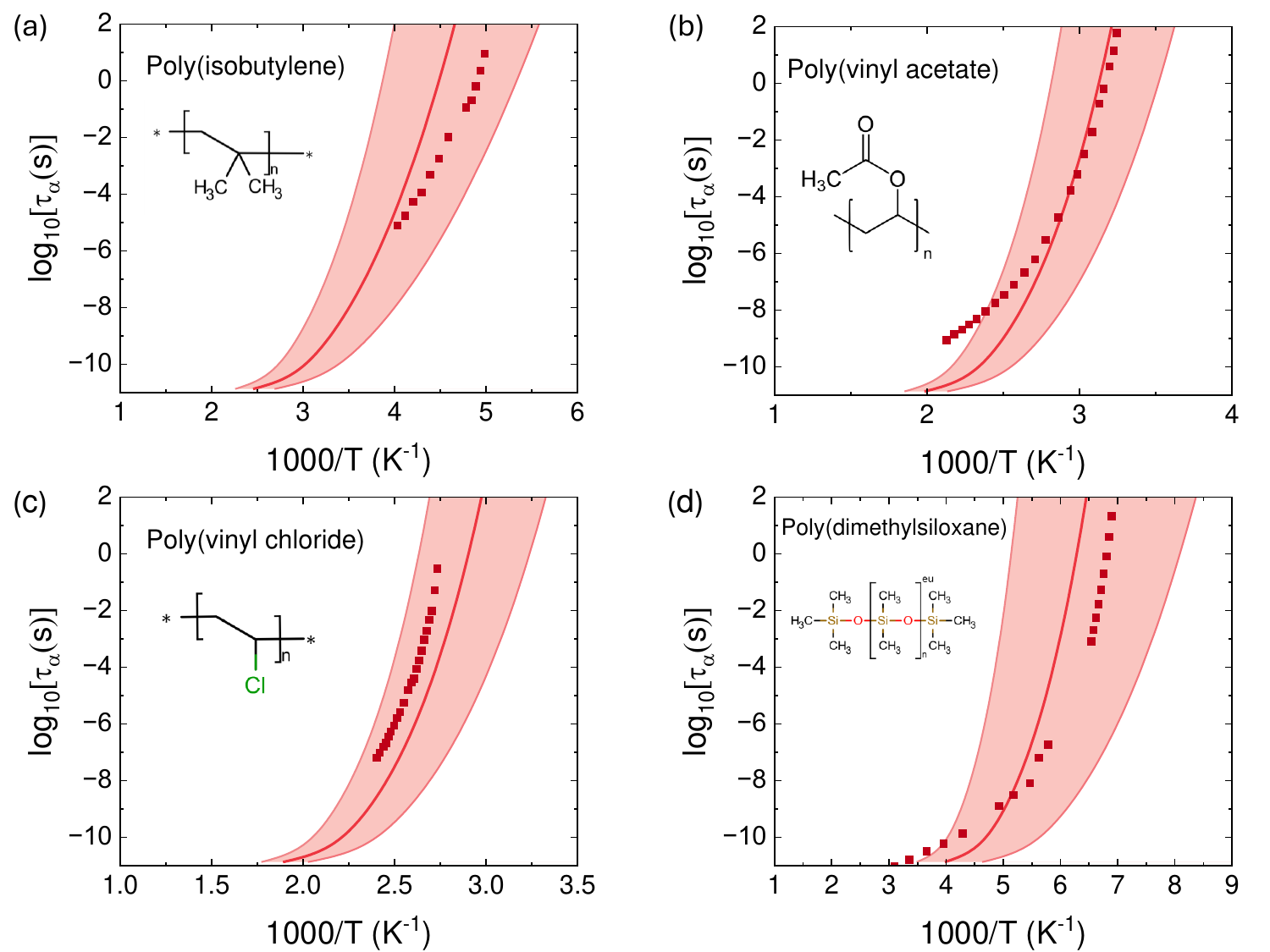}
\caption{\label{fig6}(Color online) Structural relaxation times plotted as a function of inverse temperature (1000/$T$) for four representative polymers: (a) poly(isobutylene) (PIB) \cite{Tg_relaxation_PIB}, (b) poly(vinyl acetate) (PVAc) \cite{Tg_relaxation_PVAc}, (c) poly(vinyl chloride) (PVC) \cite{Tg_relaxation_PVC} , and (d) poly(dimethylsiloxane) \cite{Tg1}. Data points correspond to experimental data. The three solid curves correspond to ECNLE calculations incorporating $T_g$ values predicted by the GPR model. The central curve is the ECNLE prediction using the true GPR-predicted $T_g$, while the two adjacent curves correspond to ECNLE results based on $T_g$ values adjusted by $\pm10 \%$ to account for the typical relative uncertainty associated with GPR predictions.}
\end{figure*}

Crucially, unlike earlier studies \cite{ML0, ML1, ML2, ML3, ML4, ML5, ML6, ML7, ML8, ML9, 50} that focused exclusively on predictive accuracy, we incorporate machine-learning-derived $T_g$ values into the ECNLE theory to compute the temperature dependence of the structural relaxation time. Figure~\ref{fig6} presents the theoretical and experimental structural relaxation times as a function of inverse temperature for four representative polymers. It is important to note that PIB, PVAc, PVC, and PDMS are not included in the training or testing dataset used to develop the machine learning models. In each case, ECNLE predictions based on $T_g$ values obtained from Gaussian Process Regression are compared with experimental BDS data. The central solid curve in each subfigure corresponds to ECNLE calculations using the true predicted GPR $T_g$, while the adjacent curves are ECNLE predictions incorporating a relative uncertainty $\pm10\%$ in $T_g$. Across all four polymers, the agreement between the experimental data and ECNLE predictions is excellent. Remarkably, most experimental data points fall within the confidence interval (shaded band in Figure~\ref{fig5}). These findings clearly indicate that the GPR-predicted values $T_g$ integrated into the ECNLE theory can quantitatively accurately predict the temperature-dependent relaxation behavior in polymers. They further validate the effectiveness of combining machine learning with physics-based theory to achieve reliable and predictive modeling. Our combined framework not only preserves predictive accuracy but also provides a powerful alternative to relying solely on experimentally measured $T_g$ values, particularly for polymers that are computationally designed or experimentally inaccessible.

\begin{figure*}[htp]
\includegraphics[width=16cm]{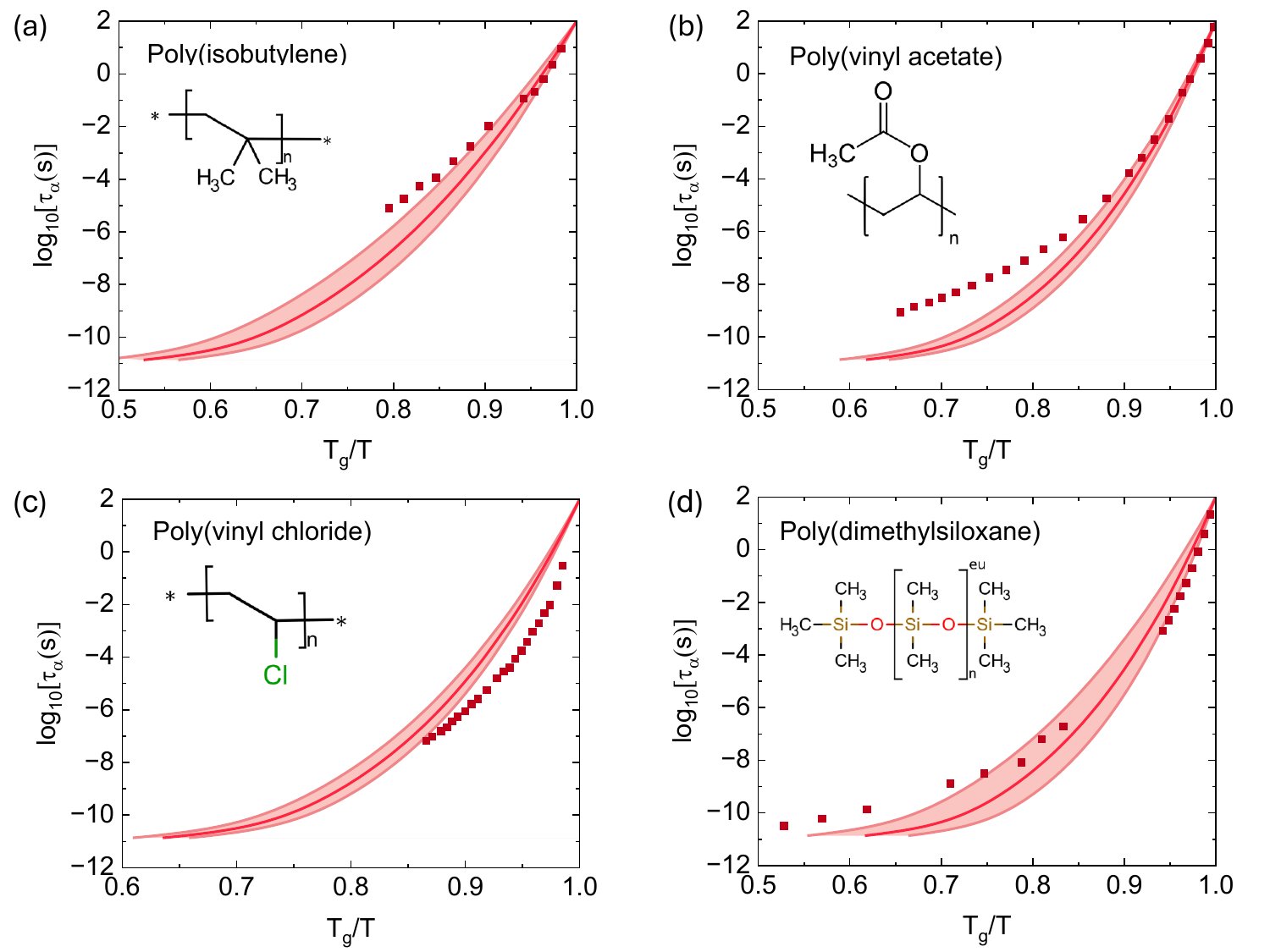}
\caption{\label{fig7}(Color online) Structural relaxation times plotted as a function of $T_g/T$) for (a) PIB\cite{Tg_relaxation_PIB}, (b) PVAc \cite{Tg_relaxation_PVAc}, (c) PVC \cite{Tg_relaxation_PVC}, and (d) PDMS \cite{Tg1}. Data points correspond to experimental data. Solid curves represent ECNLE calculations as shown in Fig. \ref{fig6} using GPR-predicted $T_g$ values.}
\end{figure*}

To further evaluate the accuracy of our theoretical framework, we replot the data of Fig. \ref{fig6} in the Angell representation and show results in Fig. \ref{fig7}. This representation facilitates a direct comparison of the temperature-dependent relaxation dynamics and dynamic fragility with previously reported values in Ref. \cite{20, 31}. From the calculations, the estimated dynamic fragilities for PIB, PVAc, PDMS, and PVC are approximately 61, 81, 80, and 88, respectively. These values are in reasonable agreement with experimental data reported in Refs.\cite{20,31}, which are 46, 95, 79–100, and 191 for PIB, PVAc, PDMS, and PVC, respectively. While our calculations slightly underestimate the fragility in some cases, the deviations are acceptable, especially considering that our model does not rely on any adjustable parameters. Notably, the ECNLE studies in Ref. \cite{20} in conjunction with the $S_0(T)$-based thermal mapping predicted fragilities of 85 for PIB, 95 for PVAc, 84 for PDMS, and 94 for PVC. Although this approach includes more detailed thermodynamic input, its predictions still show noticeable deviations. The extended ECNLE theory in Ref.\cite{31} further improves agreement by introducing a material-specific parameter $a_c$, this enhancement comes at the cost of reduced generality. Since our model avoids such parameters, direct comparison should be interpreted with caution. For PVC, in particular, the fragility reported in Refs.~\cite{20,31} deviates significantly from the experimental value and reveals that even more sophisticated models can face limitations for certain systems. Overall, our simplified ECNLE framework remains effective in capturing the fragility trends across diverse polymers.

Based on the calculations and analysis presented above, we now establish a practical workflow for investigating the glassy dynamics of new or unsynthesized polymers. The process begins by providing the SMILES representation of a polymer as input to the trained Gaussian Process Regression model to predict the glass transition temperature. This predicted $T_g$ is subsequently incorporated into the ECNLE theory to calculate the temperature dependence of the structural relaxation time. Specifically, the $T_g$ value is used in the thermal mapping relation defined in Eq.~(\ref{eq:5}). To account for the distinct thermal expansion behaviors observed in polymers with different $T_g$ regimes, the effective thermal expansion coefficient is assigned based on the predicted $T_g$. For polymers with $T_g > 160$ K, we adopt $\beta = 12 \times 10^{-4}$ K$^{-1}$, while for those with $T_g < 160$ K, a larger value of $\beta = 24 \times 10^{-4}$ K$^{-1}$ is used to reflect the enhanced thermal expansivity expected in low-$T_g$ systems. Although the approximation of a constant $\beta$ simplifies the modeling approach and reduces the need for extensive experimental input,future work could examine how a temperature-dependent thermal expansion coefficient might improve the accuracy of ECNLE predictions. This approach provides a data-driven pipeline for predicting the glassy dynamics of polymers.

It is worth noting that a recent work \cite{52} has shown that including thermodynamic and vibrational descriptors can enhance the ability of machine learning models to investigate glassy dynamics in materials. Although our current approach uses only chemical structure and composition-based features, it captures similar physical behavior through the use of the thermal expansion coefficient in our simplified thermal mapping. This parameter represents how the material responds to temperature changes and indirectly includes information related to molecular vibrations and local softness. The quantitative agreement between our theoretical predictions and experimental data clearly supports to the fact that thermodynamic and vibrational properties are important indicators of glassy relaxation.

\section{Conclusion}
In summary, we have developed a unified framework that integrates machine learning with the ECNLE theory to enhance the predictive modeling of glassy dynamics in polymers. By training and validating several regression models on a large chemically diverse dataset of 7174 polymers, we have identified Gaussian Process Regression as the most accurate and generalizable model for predicting glass transition temperatures. Its performance is better than most previously reported approaches in both accuracy and robustness across chemical space. Importantly, by integrating GPR-predicted $T_g$ values into the ECNLE theory coupled with a simplified thermal mapping (Eq. (\ref{eq:5})), we successfully captured the temperature dependence of structural relaxation times across various polymers. The theoretical predictions show excellent agreement with experimental BDS data. For polymers with low predicted $T_g$ values (below 160 K), we increase the thermal expansion coefficient $\beta$ in the mapping equation from $12 \times 10^{-4}$ K$^{-1}$ to $24 \times 10^{-4}$ K$^{-1}$. This modification significantly improves the accuracy of the predicted relaxation behavior and suggests the potential need for future experimental studies to validate thermal expansion properties in low-$T_g$ materials.

\section*{Supplementary Materials}
The supplementary materials include detailed benchmarking results for machine learning models trained to predict the glass transition temperature of polymers with performance metrics for different train-test splits. It also presents extended ECNLE calculations using chemically specific polymer models based on PRISM theory and compares them with results from the hard-sphere fluid model to validate the simplified approach used in the main text.

\begin{acknowledgments}
This research was funded by the Vietnam National Foundation for Science and Technology Development (NAFOSTED) under Grant No. 103.01-2023.62.
\end{acknowledgments}
\section*{Conflicts of interest}
The authors have no conflicts to disclose.

\section*{Data availability}
The data that support the findings of this study are available from the corresponding author upon reasonable request.

\end{document}